\documentclass[prc,nofootinbib,showpacs,twocolumn]{revtex4}
\usepackage{graphicx}
\usepackage[usenames,dvipsnames]{color}
\usepackage{amsmath,amssymb,bbold}
\usepackage{hyperref}
\begin{document}
\title{Towards laboratory detection of topological vortices in 
superfluid phases of QCD}  
%
%
\author{Arpan Das$^{1,2}$ \footnote {email: arpan@iopb.res.in}}
\author{Shreyansh S. Dave$^{1,2}$ \footnote {email: shreyansh@iopb.res.in}}
\author{Somnath De$^{1,2}$ \footnote {email: somnath.de@iopb.res.in}}
\author{Ajit M. Srivastava$^{1,2}$ \footnote{email: ajit@iopb.res.in}}
\affiliation{$^1$ Institute of Physics, Bhubaneswar 751005, India\\
$^2$ Homi Bhabha National Institute, Training School Complex,
Anushakti Nagar, Mumbai 400085, India}

\begin{abstract}
Topological defects arise in a variety of systems, e.g. vortices in 
superfluid helium to cosmic strings in the 
early universe. There is an indirect evidence of neutron superfluid 
vortices from glitches in pulsars. One also expects that  
topological defects may arise in various high baryon density phases of 
quantum chromodynamics (QCD), e.g. superfluid topological vortices in 
the color flavor locked (CFL) phase. Though vastly different in 
energy/length scales, there are universal features, e.g. in the formation
of all these defects. Utilizing this universality, we investigate the 
possibility of detecting these topological superfluid vortices in 
laboratory experiments, namely heavy-ion collisions. Using hydrodynamic 
simulations, we show that vortices can qualitatively affect the power 
spectrum of flow fluctuations. This can give unambiguous signal for 
superfluid transition resulting in vortices, allowing for check of 
defect formation theories in a relativistic quantum field theory system,
and the detection of superfluid phases of QCD. Detection of nucleonic
superfluid vortices in low energy heavy-ion collisions will give
opportunity for laboratory controlled study of their properties, 
providing crucial inputs for the physics of pulsars.  
\end{abstract}

\maketitle


\section{Introduction}
Topological defects are typically associated with symmetry breaking 
phase transitions. Due to their topological nature, they display various 
universal properties, especially in their formation mechanism and 
evolution. This has led to experimental studies of defect formation in 
a range of low energy condensed matter systems, e.g., superfluid helium, 
superconductors, liquid crystals etc. \cite{zurek,jphysg} which have 
utilized this universality and have provided experimental checks on 
various aspects of the theory of cosmic defect formation, usually 
known as the Kibble mechanism \cite{kbl}. However, it is clearly desirable 
to experimentally test these theories also in a relativistic quantum field 
theory system for a more direct correspondence with the theory of cosmic 
strings and other cosmic defects. 

We address this possibility in this paper and focus on 
heavy-ion collision (HIC) experiments. One of
the main aims of these experiments is to probe the QCD phase 
diagram which shows very rich features, especially in the regime of 
high baryon density and low temperatures. FAIR and NICA are upcoming 
facilities for HIC, dedicated to the investigation 
of high baryon density phases of QCD. Exotic partonic phases 
e.g.  two flavor color superconducting (2SC) phase, crystalline
color superconducting phase, color flavor locked (CFL) phase,
\cite{cfl} etc. are possible at very high baryon density. 
Transitions to these phases is associated with complex
symmetry breaking patterns allowing for a very rich
variety of topological defects in different phases. 
Even at moderately low baryon densities, nucleon superfluidity
(neutron superfluidity and proton superconductivity) arises.
The CFL phase occurs at very high baryon densities, with baryon densities 
at least an order of magnitude higher than the nuclear saturation density
($\rho_0$), and temperatures up to 
about 50 MeV, whereas nucleonic superfluidity occurs at much lower 
densities, of order $(10^{-3} - 1) \rho_0$, and temperatures as low 
as 0.3 MeV. Interestingly, this entire vast range of densities and 
temperatures may be accessible at the facilities such a FAIR and NICA. As we 
noted above, irrespective of the energy scale, universality of defect 
formation allows us to infer reasonably model independent predictions about
qualitative effects arising from vortex formation from these
different phase transitions.

 In the present day universe, superfluid phases of nucleons are expected 
to exist inside neutron 
stars \cite{superfluid} and resulting vortices are supposed to be
responsible for the phenomenon of glitches \cite{nstar}. No such 
observational support exists yet for the high density phases of QCD 
(e.g. CFL phase) in any astrophysical object.
In an earlier paper, some of us have proposed the detection
of such phase transitions by studying density fluctuations arising
from topological defect formation and its effects on pulsar timings
and gravitational wave emission \cite{nstarflct1,nstarflct2}. 

All of the HIC investigations in the literature
probing the high baryon density regime of
QCD have focused primarily on signals related to the quark-hadron 
transition. We propose a somewhat different line of 
focus at these experiments. Some of these exotic high baryon density 
partonic phases also have superfluidity. For example, the CFL phase 
corresponds to the spontaneous symmetry breaking pattern, $SU(3)_{color} \times
SU(3)_L \times SU(3)_R \times U(1)_B \rightarrow SU(3)_{color+L+R} 
\times Z_2$.  Superfluidity arises from spontaneous breaking 
of $U(1)_B$ to $Z_2$ as the diquark condensate for the CFL 
phase is not invariant under $U(1)_B$ baryon number transformations.
This is also expected in somewhat lower density phases
(where effects of heavier strange quark become important) such as 
the CFL+$K^0$ phase \cite{cflk0}. In HIC, if any of these phases arise, a 
superfluid transition will inevitably lead to production of superfluid 
vortices via the Kibble mechanism \cite{kbl}. 

 Similarly, for relatively lower energy heavy-ion collisions, the hot
 nucleonic system formed in the collisions may undergo transition
 to nucleonic superfluid phase as it expands and cools.
This will again lead to the formation of nucleonic superfluid vortices
via the Kibble mechanism. Note, these are precisely the same vortices which
are believed to play crucial role in pulsar glitches, though there they form
due to rotation of the neutron star.  As we will discuss later, 
universality of defect formation in the Kibble mechanism tells that defect
density of order one will be produced per correlation domain \cite{kbl}.
(For a second order transition, critical slowing down can affect defect 
formation in important ways, and is described by
the Kibble-Zurek mechanism\cite{zurek}.) 

It is immediately obvious that the 
most dramatic effect of presence of any vortices will be on the resulting
flow pattern. We carry out detailed simulations of development of flow
in the presence of vortices and study qualitative changes in the flow pattern.

\section{Kibble mechanism, vortex formation and local linear momentum 
conservation}

 We briefly recall the basic physics of the Kibble mechanism which
originates from the formation of a sort of
 domain structure during a phase transition. The order parameter field
 (superfluid condensate in this case) is correlated (hence can be 
approximately taken to be uniform) inside a domain while it varies 
randomly from one
 domain to another. Such a picture of domains is very natural for a
 first order transition via bubble nucleation with each bubble being
 an independent domain. Even for a second order transition, correlation
 length size regions correspond to such domains. For a superfluid transition,
 the phase of the order parameter varies randomly from one domain
 to another (the magnitude of the order parameter being fixed by the
 temperature). As the gradient of the phase directly correspond to
 superfluid velocity, spontaneous generation of flow is inevitable
 in a phase transition. Further, at the junction of several domains
 one can find non-zero circulation of flow if the order parameter
 phase winds non-trivially around the junction. These are superfluid
 vortices. This picture of formation of vortices is actually very general
 and applies to the formation of all types of topological defects
 in symmetry breaking transitions.

 However, spontaneous formation of superfluid vortices via Kibble mechanism 
in a transition from  normal to superfluid phase has certain non-trivial 
aspects which are not present in the formation of other types of topological 
defects. During phase transition, the spontaneous generation of flow of the 
superfluid, as mentioned above, is not allowed by local linear momentum 
conservation.
Basically, some fraction of atoms (e.g. $^4$He atoms) form the
superfluid condensate during the transition and develop momentum
due to the non-zero gradient of the phase of the condensate. The only
possibility is that the remaining fraction of atoms (which form the
normal component of fluid in the two-fluid picture) develop opposite 
linear momentum so that the momentum is local conserved. This means
that even though order parameter phase gradients are present across
different domains generating superfluid flow across different
domain junctions, there is no net momentum flow anywhere in the
beginning. Note, this argument is somewhat different from the conventional
argument of angular momentum conservation for Kibble superfluid
vortices where one knows that spontaneous generation of net rotation of
the superfluid has to be counter balanced by the opposite rotation of the 
vessel containing the superfluid. Here, we are arguing for local linear 
momentum conservation.

The immediate implication of this local linear momentum conservation is
that the initial velocity profile for the normal fluid around each vortex
formed via the Kibble mechanism should be exactly the same as the velocity 
profile of the superfluid velocity profile (as determined by the
local momentum conservation at the time of vortex formation, depending
on relative fraction of the normal fluid and the superfluid).
The momentum balance is being achieved locally here, simply by
the normal component of fluid recoiling to balance the local momentum
generated for the superfluid component. So, basically, some particles
fall into a quantum state with non-zero momentum, which, for an
isolated system, is only possible when other particles in that part of 
the system develop equal and opposite momentum.
The final picture is then that, spontaneous generation of vortex
via the Kibble mechanism leading to superfluid circulation in such a
system will be accompanied by opposite circulation being generated
in the normal component of the fluid (to balance the 
momentum conservation).

 We mention here an important implication of the above discussion.
In standard application of the Kibble mechanism for superfluid $^4$He
transition one expects a dense network of superfluid vortices
which should be detectable in experiments. However, above arguments
show that at the time of formation, superflow and normal flow have
opposite flows, so experimental detection may become very complicated.
As normal flow will be expected to change in time due to viscous
effects one may expect easier detection at later times. However,
the vortex network itself evolves and coarsens rapidly in time,
thus complicating inference regarding Kibble estimate of vortex
formation. In conclusion, counter balancing normal fluid flow
which necessarily arises in Kibble mechanism must be accounted
for when comparing theoretical predictions with data.  We plan to carry 
out a detailed investigation of this issue in a future work.

\section{Hydrodynamical simulation of flow fluctuations with
vortices}

We will first focus on superfluid transitions in the high baryon 
density partonic phase of QCD and later comment on the possibility
of low baryon density nucleonic superfluid phase transition. 
We carry out hydrodynamical simulations of the evolution of a partonic
system in the presence of vortices using a two-fluid picture of superfluid.
We also consider a range of values for the density fraction of superfluid 
to  normal fluid and study its effect on the signals. The two fluids are 
evolved, as in our earlier simulations \cite{flow}, with  Woods-Saxon 
profile of energy density with and without additional density fluctuations 
(though it does not appear to have crucial effects on our results). 
It is known that various high baryon density partonic phases (QGP, 2SC, CFL 
etc.) do not differ much in energy density and pressure \cite{cfl}.
Thus, we evolve the superfluid component with the same equation of state as 
the normal fluid, which is taken simply to be an ideal 
gas of quarks and gluons at temperature $T$ and quark
chemical potential $\mu_q$ with the energy density $\epsilon$ given 
as (for two light flavors) \cite{qgp},

\begin{equation}
\epsilon = {6 \over \pi^2} \left({7\pi^4 \over 60}T^4 +
{\pi^2 \over 2}T^2\mu_q^2 + {1 \over 4}\mu_q^4 \right)
+ {8\pi^2 \over 15}T^4 
\end{equation}

with pressure $P = \epsilon/3$. Note, as our interest is only
in discussing the hydrodynamics in the partonic phase (and not in the
quark-hadron transition), we  do not include the bag constant. 
The energy-momentum tensor is taken to have the perfect fluid form,

\begin{equation}
T^{\mu\nu} = (\epsilon + P)u^\mu u^\nu - P g^{\mu\nu}
\end{equation}

where $u^\mu$ is the fluid four-velocity. The hydrodynamical evolution
is carried out using the equations,  $\partial_\mu T^{\mu\nu} = 0$.
Note that we do not need to use conservation equation for the
baryon current as our interest is only in flow pattern requiring
knowledge of $\epsilon$ and $P$ and the ideal gas equation of state 
relating $P$ and $\epsilon$ does not involve $\mu_q$.
The simulation is carried out using a 3+1 
dimensional code with leapfrog algorithm of 2nd order accuracy. For 
various simulation details  we refer to the earlier work \cite{flow}. 
The initial energy density profile for both fluid components (normal 
fluid as well as superfluid) is taken as a Woods-Saxon background 
of radius 3.0 fm with skin width of 0.3 fm (with appropriate fractions
of energy density). We take the initial central 
energy density $\epsilon_0$ with temperature $T_0$ = 25 MeV and 
$\mu_q = 500$ MeV as representative values \cite{cfl}.  Initial random 
fluctuations are incorporated in terms of 10 randomly placed Gaussian of 
half-width 0.8 fm, added to the background energy density, 
with central amplitude taken to be $0.4\epsilon_0$.

The initial velocity profile is determined by the fluid rotation
around the vortices. For the superfluid part,
The magnitude of the fluid rotational 
velocity at distance $r$ from the vortex center is taken as 

\begin{equation}
v(r) = v_0 {r \over \xi} \quad (r \le \xi);
\qquad v(r) = v_0 {\xi \over r} \quad (r > \xi)
\end{equation}

 Here $\xi$ is the coherence length. For CFL vortex, 
estimates in ref.\cite{superfluid} give $v_0 = 1/(2\mu_q \xi)$ and
the coherence length is given by

\begin{equation}
\xi \simeq 0.26 \left({100 {\rm MeV} \over T_c}\right)
\left(1-{T \over T_c}\right)^{-1/2} ~~{\rm fm}.
\end{equation}

As we mentioned above,  exactly at 
the time of formation of the vortex, the velocity profile of the normal 
component will be opposite, having exactly the same form as that of 
the superfluid vortex, but with a magnitude appropriate for the fraction
of the normal fluid. So, for the normal fluid, the initial velocity
profile is taken to be exactly the same as given by Eqn.(3), but with 
$v_0$ having opposite sign, and suitably scaled for local momentum 
conservation depending on superfluid density fraction. This will remain 
as correct profile if the normal fluid has very low viscosity (note,
QGP at RHIC energies has low viscosity).  However, if the viscosity is 
significant,  then this velocity profile will not 
be sustained due to differential
rotation, and will change in time. We have accounted for this possibility 
also by considering admixture of velocity profile for viscous fluid with a
velocity profile $v(r) \propto r$ at different  times (even though we use 
non-viscous hydrodynamics). We find that this does not affect
the qualitative features of our results at all, except that with
large fraction of this viscous velocity profile one also gets a 
non-zero directed flow in the presence of vortices.

 We take value of superfluid transition temperature $T_c = 50$ MeV
\cite{superfluid}. For the initial central temperature $T_0$ = 25 MeV, 
resulting values of $\xi = 0.7$ fm and $v_0 = 0.3$ (we take $c=1$). 
(Note, even though we use 2-flavor equation of state, we use the estimates 
of the vortex velocity profile for the CFL phase for order of magnitude 
estimates.)

 Formation of vortices in superfluid transition will be in accordance
with the Kibble mechanism as we discussed above. We will not actually simulate
the Kibble mechanism here as our interest is not in getting a statistical
network of defects. Rather, we want to see effect of a couple of
vortices on the resulting flow pattern. As we will see below, for
the size of QGP region taken here, the number of superfluid vortices
expected here is of order 1. We do not simulate coupled dynamics of
normal and superfluid components. Instead, we evolve the two 
components using separate conservation equations for the two energy 
momentum tensors. This allows us to simulate a delayed superfluid
transition. This models the situation when initial partonic system has 
too high a temperature (but with appropriate baryon density)
to be in the superfluid phase, though it is still in the 
QGP phase, and subsequent expansion and cooling leads to crossing
the phase boundary to the superfluid phase. Also, for the case of 
nucleon superfluidity (to be discussed below), initial high temperatures 
will lead to normal nucleonic phase, and only at late stages of expansion
superfluid phase may arise.  In a coupled fluid dynamics, this cannot 
be achieved as one always has a normal fluid as well as a superfluid 
component.     

 For observational signatures, we focus on the power spectrum of
flow fluctuations. In a series of papers some of us have demonstrated 
that just like the power spectrum of CMBR, in HIC also the
power spectrum of flow fluctuations has valuable information about
the initial state fluctuations of the plasma \cite{cmbhic1,cmbhic2}. We will 
calculate the power spectrum of flow fluctuations
and study the information contained in the power spectrum about
the initial vortex induced velocity fields. We focus on the central rapidity
region (focusing on a thin slab of width 2 fm in z direction at
z = 0) and study the angular anisotropy of the fractional fluctuation in
the transverse fluid momentum,  $\delta p(\phi)/p_{av}$, where $\phi$ is the 
azimuthal angle, $p_{av}$ is the angular average of the transverse fluid 
momentum, and $\delta p(\phi) = p(\phi) - p_{av}$. 
This fluid momentum anisotropy is eventually 
observed as momentum anisotropy of the hadrons which are finally detected.
The power spectrum of flow fluctuations is obtained by calculating
the root mean square values $v_n^{rms}$ of the $n_{th}$ Fourier 
coefficient $v_n$ of the  momentum anisotropy ${\delta p(\phi)}/p_{av}$. 
We use lab fixed coordinates, so event averaged value of $v_n$ is zero.

  We use standard Kibble mechanism, as described above, to estimate the 
probability of vortex formation. In the CFL phase, superfluidity corresponds
to spontaneous breaking of U(1) symmetry (just like the case
for superfluid $^4$He, though for $^4$He case U(1) is completely
broken while for the CFL phase, U(1) breaks to $Z_2$). 
In two space dimensions, this leads to 
probability 1/4 for the formation of a vortex (V) or antivortex (AV) per 
correlation domain \cite{kbl}. For the azimuthal momentum anisotropy
in the central rapidity region, the relevant velocity
field is  essentially two-dimensional. With the correlation length of
order 1 fm, and the plasma region which we are taking to have
a radius of 3 fm, we expect number of superfluid vortices to be  about 2. 
For definiteness, we will consider cases
of 1 vortex, and a V-V pair and a V-AV
pair. The locations of these are taken to be randomly distributed
in the plasma region. To have clear signals, we have taken 
definite orientations for the vortices. We consider vortices
either pointing along z axis (with random locations) or
pointing along x axis (passing through the origin).

\section {Results of the simulation}

 We now present results of the simulations. Fig.1 shows the effect of 
 vortices on the flow power spectrum for a central collision at 
 $\tau-\tau_0 = 1.68$ fm, (with $\tau_0 = 1.0$ fm). We mention that 
 with our numerical code, fluid evolution becomes unstable for large 
times, especially with complex flow pattern with high velocities,
hence we show the results at relatively shorter times. However,
these qualitative signals will be expected to survive even for longer 
times, though with possibly smaller magnitudes. As such these will apply to 
situations of early freezeout, e.g.  for smaller nuclei, or for peripheral 
collisions. Fig.1 shows plots of $v_n^{rms}$ for the cases of no vortex, 
one vortex, a V-V pair, and a V-AV pair. In all cases, vortices are taken
along the z axis with random positions. Noteworthy is a large value
of the elliptic flow for the V-AV case (even though
this is a central collision). For all cases with vortices we find
that the elliptic flow is very large initially (see, also, Fig.2). This
is clearly seen in the inset of Fig.1 for the V-AV case which also shows 
the dependence of elliptic flow on superfluid fraction and its time 
evolution. This can be detected by its effects on photon or dilepton 
elliptic flow \cite{dks} which is sensitive to flow effects at very early
stages.

\begin{figure}[ht!]
\includegraphics[width=6cm,angle=270]{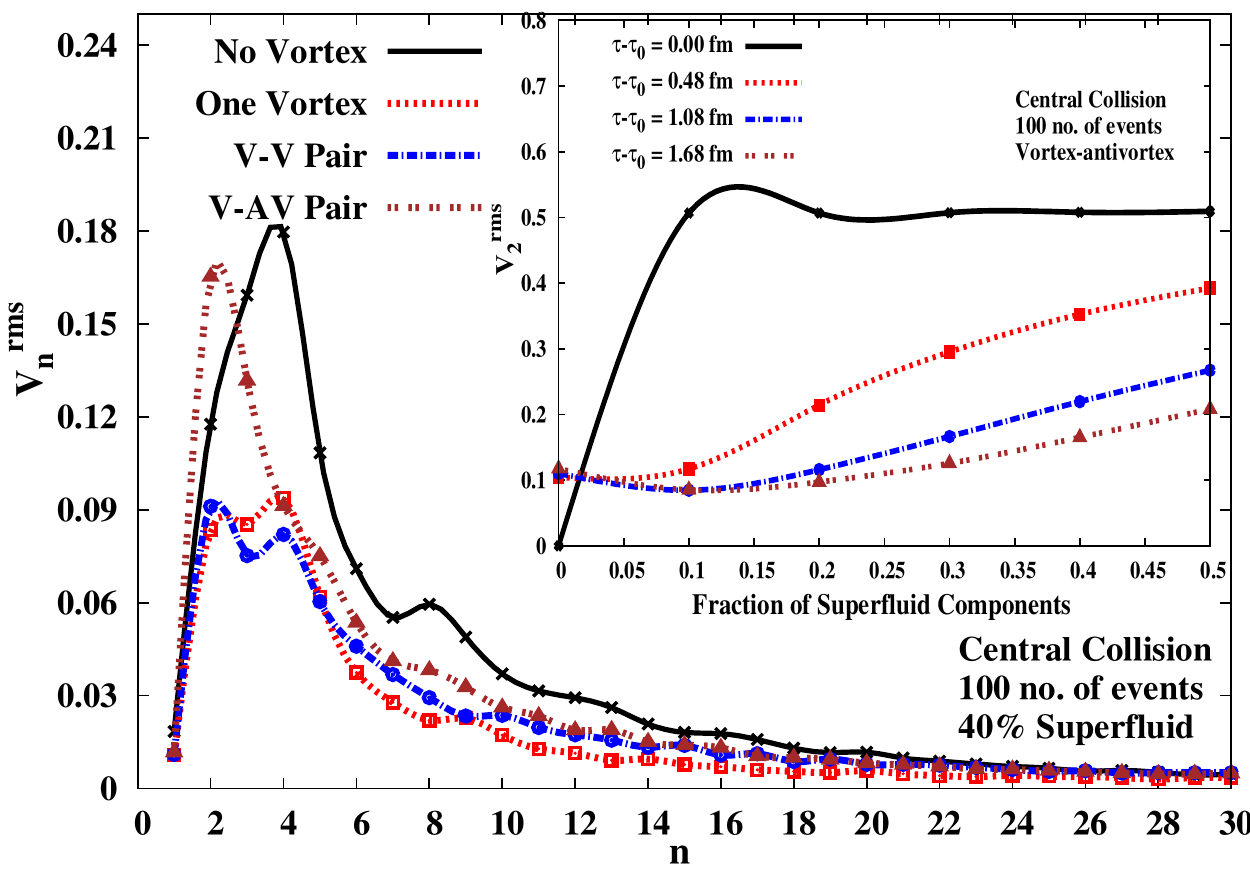}
\leavevmode
\caption{Power spectrum at $\tau-\tau_0 = 1.68$ fm for central collision.
Different plots show the power spectrum for the cases of no vortex,
single vortex, a V-V pair, and a V-AV pair. Inset shows 
dependence of elliptic flow at different times on the superfluid 
fraction for the V-AV case showing very large initial elliptic flow.} 
\end{figure}

Fig.2 shows the time evolution of the power spectrum for the case with a 
V-V pair (we find similar results for V-AV case as 
well). Note difference in the power for
even and odd Fourier coefficients at earlier times. (Such a qualitatively
different pattern in HIC has only been predicted in
the presence of strong magnetic field, as reported in ref. \cite{bfield}).
This result also has 
interesting implications for the CMBR power spectrum. It is known
that low $l$ modes of CMBR power spectrum also show difference
in even-odd modes \cite{evenodd}. It is possible that this feature
may be indicative of the presence of a magnetic field, or presence of
some vorticity  during the very early stages of the inflation.

\begin{figure}[ht!]
\includegraphics[width=8cm]{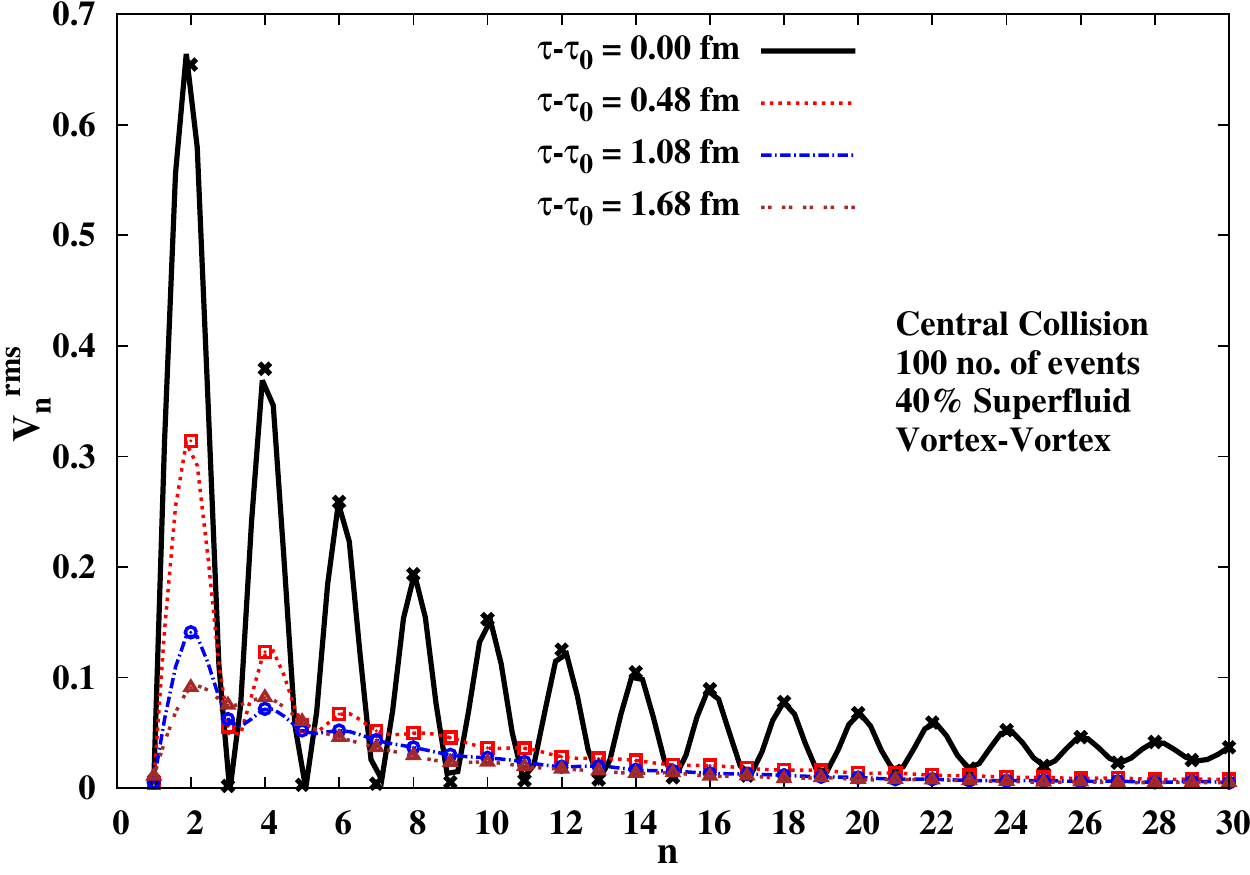}
\leavevmode
\caption{Time evolution of the power spectrum for the case with a 
V-V pair showing the difference in the power 
for even and odd Fourier coefficients for early times.}
\end{figure}

 Fig.3 presents the case of non-central collisions. Here we 
consider an ellipsoidal shape for the plasma region as appropriate for
a non-central collision with semi-minor axis along the x-axis, and 
initial spatial eccentricity = 0.6. Here we plot $v_2$ for a single
event (not the rms value), for two cases, a V-AV pair 
pointing in z direction and located on the x-axis at $x = \pm 1.5 
fm$ respectively, and the other case with a single vortex lying along 
the x-axis. Both cases show strongly negative elliptic flow at initial
stages. Fig.3 also shows large (negative) values of $v_4$ for both
these cases which arises from vortex induced elliptic flow being
in the orthogonal direction to the shape induced elliptic flow.
These large values of negative elliptic flow as well as $v_4$ may be 
observed  if the freezeout occurs at early times (in smaller systems, 
or in peripheral collisions) and should also leave imprints on other 
observables such as on
$v_2$ for photons \cite{dks}. Note that negative elliptic flow
can arise in relatively low energy HIC  due to
squeeze-out effects \cite{fope}. However, for low energy 
collisions (as we discuss below for nucleonic superfluidity),
a vortex induced negative elliptic flow is completely
uncorrelated to the elliptic shape of the event (which can be
inferred from independent observables), hence can be distinguished
from the squeeze-out effect. Further , at higher 
energies (where CFL phase may be expect to arise), no
squeeze-out is expected, so a negative elliptic flow can 
signal vortex formation. 

We have also carried out all the simulations 
with a delay of up to 1 fm in the onset of superfluid transition 
(following our  modeling of the two fluid picture as explained above). 
The results remain essentially unchanged with various plots 
showing changes of order only few percent. 

\vskip .2cm
\begin{figure}[ht!]
\includegraphics[width=8cm]{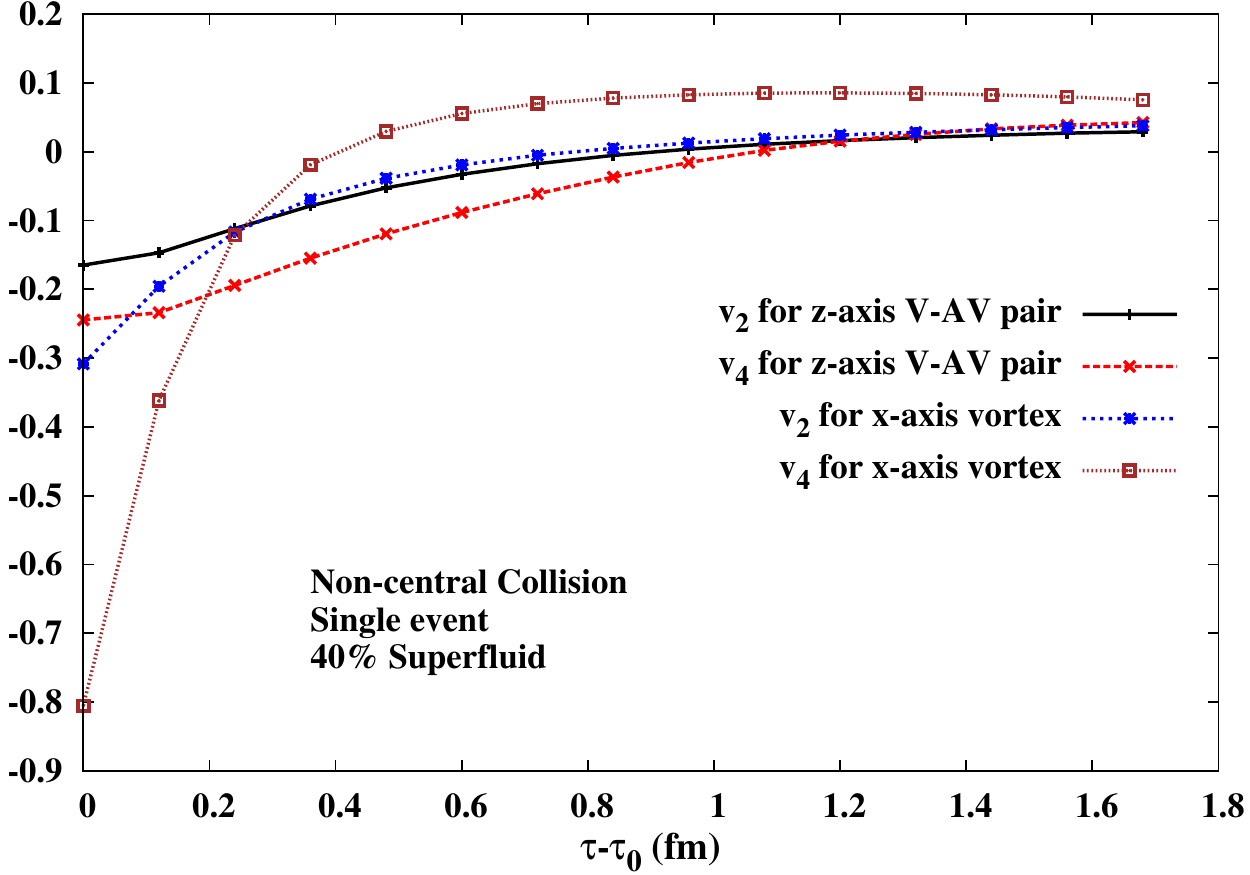}
\leavevmode
\caption{Plot of $v_2$ and $v_4$ for non-central collisions for a
V-AV pair along z axis, and  a single vortex 
along the x axis, showing negative elliptic flow at initial
stages as well as large (negative) values of $v_4$.}
\end{figure}

\section{Nucleonic superfluidity for low energy collisions}

We now discuss the possibility of detecting nucleonic superfluidity
in HIC. Though neutron superfluid condensate is expected to
exist inside several nuclei, these systems are typically too small to 
demonstrate bulk superfluid phase and its associated superfluid 
vortices, as are expected inside a neutron star.
Calculations for neutron stars show that
nucleonic superfluidity is expected in range of densities from
10$^{-3} \rho_0$ (for $^1S_0$ pairing of neutrons) to few times
$\rho_0$ (for $^3P_2 - ^3F_2$ pairing). The critical temperature
can range from 0.2 MeV to 5 MeV (depending on the nuclear potential
used \cite{nstarsf1,nstarsf2}). Temperatures and densities of this order are
easily reached in HIC at relatively low energies. For example, at the 
FOPI-facility at GSI Darmstadt, temperatures of about 17 MeV 
(with $\rho \sim 0.4 \rho_0$) were reported in Au-Au collisions
at 150 MeV/nucleon lab energy \cite{T17}. Temperatures of order
4-5 MeV were reported in Au-Au collisions at E/A = 50 MeV, at heavy-ion
synchrotron SIS \cite{T5}. Thus temperatures/densities appropriate
for the transition to the nucleonic superfluid phase can easily
be reached in HIC.  Universality of defect formation
implies that the qualitative aspects of our results in this paper (for 
the CFL phase) will continue to hold even in this lower density regime.
FAIR and NICA are ideal facilities
for probing even this low energy regime with detectors suitable for
measurements with which flow power spectrum analysis can be performed.
Detection of signals as discussed in this paper can provide a clean
detection of nucleonic superfluid vortices. It is worth emphasizing
the importance of focused experiments for creating a nucleonic system
of several fm size which can accommodate nucleonic superfluid vortices.
Direct experimental evidence of these vortices and controlled studies
of their properties can provide a firm basis for our understanding
of neutron stars. This is all the more important in view of the fact
that gravitational waves from rotating neutron stars and their collisions 
will be thoroughly probed by LIGO and upcoming gravitational wave
detectors.

\section{Conclusions}

  We conclude by pointing out the importance of searching for the 
superfluid vortices during transition to high baryon density QCD phases, or 
to nucleonic superfluid phase, at FAIR and NICA. Due to universal
features of vortex (topological defect) formation, these vortices
directly probe the symmetry breaking pattern of the phase transition
providing very useful information about the QCD phase diagram.
Various high density phases of QCD such as CFL phase etc. are associated
with definite symmetry breaking patterns leading to different topological 
defects. Detection of defects thus directly probes precise nature
of symmetry breaking transition occurring in the system. In this sense,
this technique has advantage over other observational signatures which
depend on equation of state etc. as those quantities can be strongly
model dependent (in contrast to the symmetry patterns which are the
most universal features of any phase transition). In this context we mention
that there has been study of stability of CFL vortices etc. and it is
found that for certain parameter range these vortices may be 
unstable \cite{unstbl}.
Even for the unstable case, typical decay time for the vortices will be
expected to be at least of order few fm which, though very short time
for astrophysical relevance, should be long enough time for these
vortices to leave their observational signature in heavy-ion collisions.

It is hard to overemphasize the importance of detecting nucleonic
superfluid phase and associated vortices in these experiments which have
capability of providing a controlled experimental investigation of
the properties of these vortices and associated phases. Till date, there
is no direct experimental observation of nucleonic superfluid vortices,
though they provide probably the most accurate explanations of pulsar
glitches. Thus detection of these in laboratory experiments will strengthen
our understanding of pulsar dynamics. The signals we have discussed
  show qualitatively new features in flow anisotropies
  signaling the presence of vortices and the underlying superfluid
  phase in the evolving plasma. These qualitative features
  are expected to be almost model independent, solely arising from
  the vortex velocity fields. We mention that one has to properly
  account for the effects due to jets, resonance decays etc.
 to properly account for genuine hydrodynamic flow fluctuations.
 We hope to address these issues in a future work. Also, we have not
 included error bars in our plots to avoid overcrowding of the plots.
 The number of events was chosen suitably large (100 events) so that 
 the main qualitative features of the plot are above any statistical 
 fluctuations. (Our focus is mainly on the qualitative patterns of the 
 plots, in the spirit of the universal features of topological vortices
 forming at varying energy scales, and not on precise numerical value.)
 As we mentioned, due to universality of defect formation, similar 
signals are expected from  nucleonic superfluid vortices which can 
arise in low energy HIC  providing direct experimental access to the physics of
  pulsars. 

\section*{Acknowledgment}
 We are very grateful to Rajeev Bhalerao, Partha Bagchi, Srikumar Sengupta,
Biswanath Layek, and Pranati Rath for useful discussions. Some of the results 
here were presented by AMS at the conference {\it Hadronic
Matter under Extreme Conditions} held at JINR, Dubna, in Nov. 2016,
and by SSD in the conference {\it ATHIC 2016} at New Delhi, India.
We thank the participants in these meetings for very useful
comments and suggestions.


\end{document}